\documentclass[letter]{aa} % for the letters 
\usepackage{dcolumn}

\usepackage{graphicx}
\usepackage{txfonts}
\makeatletter
 \def\hlinewd#1{%
   \noalign{\ifnum0=`}\fi\hrule \@height #1 \futurelet
    \reserved@a\@xhline}
\makeatother

\newcommand{\kms}{km\,s$^{-1}$}
\usepackage{color}

\begin{document} 

    \title{Proving strong magnetic fields near to
% a magnetosphere  around
 the central black hole in
      the quasar PG0043+039 via cyclotron lines  
%\thanks{Based on observations obtained with the XMM, HST, SALT,
%and Hobby-Eberly Telescope.}
%, which is a joint project of the University
%               of Texas at Austin, the Pennsylvania State
%           University, Stanford University, Ludwig-Maximilians-Universit\"at
%            M\"unchen, and Georg-August-Universit\"at G\"ottingen.}
}

  \author{W. Kollatschny \inst{1}, 
           N. Schartel \inst{2},
           M. Zetzl \inst{1}, 
           M. Santos-Lle\'{o} \inst{2},
           P. M. Rodr\'{i}guez-Pascual  \inst{2},
           L. Ballo \inst{3},
          }

   \institute{Institut f\"ur Astrophysik, Universit\"at G\"ottingen,
              Friedrich-Hund Platz 1, D-37077 G\"ottingen, Germany\\
              \email{wkollat@astro.physik.uni-goettingen.de} 
            \and
                XMM-Newton Science Operations Centre, ESA, Villafranca del Castillo, Apartado 78, 
              28691 Villanueva de la Ca{\~nada}, Spain 
             \and 
         Osservatorio Astronomico di Brera (INAF), via Brera 28, I-20121 Milano, Italy
}

   \date{Received , 2015; accepted , 2015}
  \authorrunning{Kollatschny et al.}
   \titlerunning{Cyclotron lines in the UV spectra of
%UV/optical spectra of X-ray weak quasar
 PG0043+039}

% \abstract{}{}{}{}{} 
% 5 {} token are mandatory
 
  \abstract
  % context heading (optional)
   {
The optical luminous quasar PG0043+039 
has not been detected before in deep
   X-ray observations indicating 
the most extreme optical-to-X-ray slope index
   $\alpha_{ox}$ of all quasars.}
  % aims heading (mandatory)
   {
This study aims to detect PG0043+039 in a deep X-ray exposure.
Furthermore, we wanted to check out whether this object shows
specific spectral properties in other frequency bands.
}
  % methods heading (mandatory)
   {We took deep X-ray (XMM-Newton), far-ultraviolet (HST),
 and optical
(HET, SALT telescopes) spectra of PG0043+039 simultaneously in July 2013.}  
  % results heading (mandatory)
   {We just detected PG0043+039 in our deep X-ray exposure.
The steep $\alpha_{ox}=-2.37 \pm 0.05$ gradient 
is consistent with an unusual steep gradient
% $\alpha_{ox}=-2.37 \pm 0.05$
 F$_\nu$~$\sim$~$\nu^{\alpha}$ with
$\alpha=-2.67 \pm 0.02$
seen in the UV/far-UV continuum.
The optical/UV continuum flux has a clear maximum near 2500 \AA{}.
%together with the unusual steep spectral index towards the
%UV/far-UV spectral range.
 The UV spectrum  
is very peculiar because it shows broad humps in addition to known emission lines.
A modeling of these observed humps with cyclotron lines can explain
their wavelength positions, their relative distances, 
 and their relative intensities.
We derive plasma
temperatures of T $\sim$ 3~keV
and magnetic field strengths of  B $\sim$ 2 $\times10^{8}$ G
for the line-emitting regions close to the black hole.
}
{}

\keywords {Galaxies: active --
                Galaxies: quasars  --
%                Galaxies: nuclei  --
                Galaxies: individual:  PG0043+039   
               }

   \maketitle
%
%________________________________________________________________
%
\section{Introduction}
Active galactic nuclei (AGNs) emit enormous luminosities at all frequency
ranges from the radio to the X-ray regime.
%The observed hard X-ray emission is thought to be produced in a corona of
%hot electrons via inverse Compton scattering of the inner photons
%(Haardt \& Maraschi \citealt{haardt91}).
The overall spectral energy distribution (SED) of these broad line emitting AGN
shows general trends seen in their mean SED
(Richards et al.\citealt{richards06}).
However, some outliers exist, such as
PG0043+039.
% is the most extreme X-ray weak quasar known to date.
It is the only quasar in the PG sample 
(Schmidt \& Green,\citealt{schmidt83})
that was not detected in a dedicated deep XMM-Newton pointing
(Czerny et al. \citealt{czerny08}).
 It shows the most extreme optical to X-ray slope of all
AGNs in the Brandt et al. \cite{brandt00} sample.
PG0043+039 (z=0.38512)
 has been classified 
 as a broad absorption line (BAL) quasar (Turnshek et al.\citealt{turnshek94})
based on a broad CIV absorption.
% at a blueshift of $\sim$ 10\,000 km $s^{-1}$.
The majority of broad absorption line quasars are X-ray weak. This
is usually explained by the absorption of the outflowing wind
in combination with the wind's velocity shear.
A conclusive interpretation of the X-ray weakness in PG0043+039
 was hampered by the absence of simultaneous
measurements, which is mandatory because both the X-ray flux and the broad
absorption system are known to be variable. 
% It is so far unknown what causes
%the extreme X-ray weakness of this object.
We took simultaneous observations in the optical, UV, and X-ray of
PG0043+039 to pin down the optical to X-ray slope. Furthermore,
we wanted to test whether PG0043+039 shows additional special
spectral properties.
\section{Observations and data reduction}
We executed
simultaneous multifrequency observations of PG0043+039
in July 2013
with the original idea of probing the nature and understanding the origin
of the X-ray weakness of PG0043+039, while ensuring that we are rid of the
potential uncertainties that variability could create.
We used the XMM-Newton satellite in the medium X-ray regime, the
Hubble Space Telescope (HST) in the far-UV, and the ground-based
10m Hobby-Eberly and Southern African Large Telescopes
(HET, SALT)
in the optical to investigate this quasar in more detail. Furthermore,
this object has simultaneously been observed with the NuSTAR satellite
in the hard X-ray range (Luo et al.\citealt{luo14}).\\
%
%
%\subsection{XMM-Newton observations}
%
%PG 0043+039 was observed twice with XMM-Newton (Jansen et al. \cite{Jansen2001})
%The first observation
%was performed on June 15th 2005, under ObsId 300890101. Visual inspection
%of the EPIC images (Turner et al. \cite{Turner2001})
% did not reveal an X-ray counterpart for PG 0043+039
%and an upper limit has been derived for the source flux of
%$<$8.6~$\times$~10$^{-16}$~ergs~s$^{-1}$~cm$^{-2}$
 %(Czerny et al.\cite{czerny08}).
%The second observation was performed
%
{\bf XMM-Newton observations:}\\
PG0043+039 was observed
 on July 18, 2013 under ObsId 690830201.
We screened for low background periods (Schartel et al.\citealt{Schartel2007},
 and Piconcelli et al.\citealt{Piconcelli2005})
 and obtained clean exposure times
of 14.5~ks for pn, 29.0~ks for MOS 1, and 31.3~ks for MOS 2. PG 0043+039 is
clearly visible as a weak point source in the images of all three cameras.
%We extracted the source counts in a circle with a radius of 10 arcsec for
%pn ((Str{\"u}der et al., \cite{Strueder2001})
% and 12 arcsec for MOS (Turner et al. \cite{Turner2001})
% and,
Since the source is weak, we added the pn, MOS 1,
and MOS 2 spectra. We obtained a count rate of
1.42$\pm$0.17~$\times$~10$^{-3}$~counts~s$^{-1}$ 
for the added signal.
An intrinsically absorbed power law can describe the spectrum
(C~$=$~55.52, d.o.f.~$=$~57)
 and reveals
N$_H$~$=$~4.9$_{-3.6}^{+6.4}$~$\times$10$^{21}$~cm$^{-2}$
%N$_H$~$=$~4.86$_{-3.64}^{+6.35}$~$\times$10$^{21}$~cm$^{-2}$
and
$\Gamma$~$=$~1.55$_{-0.42}^{+0.50}$.
    Within its error margin, the power-law index
agrees with the indices commonly found for optically selected quasars
(Piconcelli et al.\citealt{Piconcelli2005}).
The intrinsic absorbing column density cannot explain the X-ray weakness of
the quasar. We obtain a flux
F(2.0--10.0~keV)~$=$~1.80$_{-0.29}^{+0.24}$~$\times$10$^{-14}$~ergs~cm$^{2}$~s$^{
-1}$
where the errors are provided for the 68\% confidence.\\
  \begin{figure*}
    \includegraphics[width=9.7cm,angle=270]{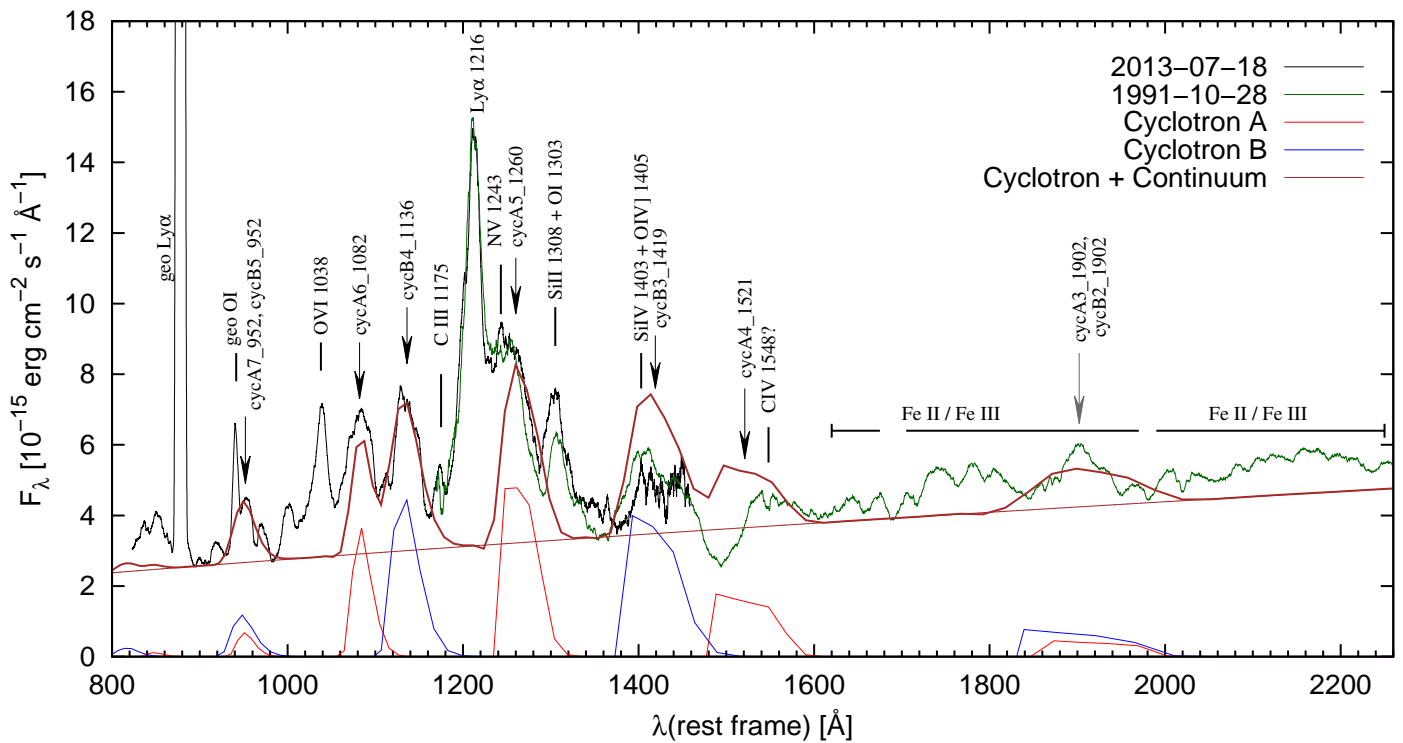}
      \caption{Combined ultraviolet spectra of PG0043+039 taken with the HST in
the years 2013 and 1991.
% We multiplied the UV spectral flux taken in 1991 with the factor 1.8
%to match the UV observations taken in 2013.
 Indicated are the identifications of the strongest UV emission lines and
of the geo-coronal lines, as well as of both cyclotron systems B and A with
their second to seventh harmonics. The integer numbers identify the
emission humps with multiples of the cyclotron fundamental.
Indicated is the power-law continuum F$_\lambda$~$\sim$~$\lambda^{\alpha}$ with
$\alpha = 0.67 \pm 0.02.$
The modeling of both cyclotron line systems A and B is given at the bottom.
}
       \vspace*{-4mm} 
         \label{pg0043_cyc_lambda.ps}
   \end{figure*}

{\bf HST-COS far-UV spectroscopy:}\\
We observed PG0043+039 over one full HST orbit with an exposure time 
of 1552 s on July 18, 2013. 
We used the COS/NUV spectrograph with the G140L grating and an 2.5 arcsec
aperture (circular diameter).
 This spectral set covers the wavelength range from 1110\,\AA\
to 2150~\AA\
with a resolving power of 2000 at 1500\,\AA{}.
For filling up the wavelength hole produced by the chip gap and for reducing
the fixed pattern noise, we split our observation into four separate segments
of 388 s duration at two different FP-POS offset positions and four different
central wavelengths.
%The observed spectrum corresponds to 830\,\AA{} to 1550~\AA\
%in the rest frame of the galaxy.
The original data were processed using the
standard CALCOS calibration pipeline.\\ 
{\bf Ground-based spectroscopy with the SALT/HET telescopes:}\\
One optical spectrum of PG0043+039 has been obtained with the
10m Southern African Large Telescope (SALT)
nearly simultaneously to the XMM/HST observations on July 21,
2013, under photometric conditions.
The spectrum was taken with the Robert Stobie Spectrograph
%(RSS; see Burgh et al., \cite{burgh03})
 attached to the telescope.
The exposure time of our spectrum amounted to 2200 s.
We covered the wavelength range from
6445 to 9400~\AA\  at a spectral resolution of 4.8~\AA\ (FWHM).
The observed wavelength range corresponds to 
a wavelength range from 4653 to 6786~\AA\ in the rest frame of the galaxy. 
A second optical spectrum of PG0043+039 was taken by us
 with the 9.2m Hobby-Eberly Telescope  (HET) at McDonald Observatory 
 on August\,1, 2013, under nearly photometric conditions.
Here we choose an exposure time of 1500 s.
We used the
Marcario Low Resolution Spectrograph (LRS)
at the prime focus of the telescope.
This spectrum covers the wavelength range from 4390\,\AA\
to 7275~\AA{}
corresponding to 3170 to 5250~\AA{}
in the rest frame of the galaxy.

The reduction of the spectra (bias subtraction, cosmic ray correction,
flat-field correction, 2D-wavelength calibration, night sky subtraction,
and flux calibration) was done in a homogeneous way with IRAF reduction
packages (Kollatschny et al.\citealt{kollatschny01}).
Details of the observations and their reduction will be given in a separated
paper (Kollatschny et al.\citealt{kollatschny15b}).
All wavelengths were converted to the rest frame of the galaxy with a redshift
of z=0.38512. 

\section{Results and discussion}
%
%\subsection{X-ray to optical spectra in PG0043+039}
%
%
\subsection{UV/far-UV and optical spectra of PG0043+039}
%\subsubsection{UV/FUV spectrum PG0043+039}
%
The UV spectrum of PG0043+039 that was taken with the
HST-FOS in July 2013 is shown in
Fig.\,\ref{pg0043_cyc_lambda.ps}.
This spectrum covers the intrinsic wavelength range of
$\sim$820\,$\AA\ $ to $\sim$1550~$\AA${}.
%The observed wavelength range from $\sim$1140\,$\AA\ $
%to $\sim$2150~$\AA\ $ corresponds to an intrinsic wavelength range of
%$\sim$820\,$\AA\ $ to $\sim$1550~$\AA\ $.
%
%
%
%                                                
%   \begin{figure*}
%    \includegraphics[bb=40 90 380 700,width=9.12cm,angle=-90]{2.25_avg_rms.eps}
%    \includegraphics[width=10cm,angle=270]{pg0043_hst.ps}
%      \caption{HST-COS FUV spectrum of PG0043+039 corrected for Galactic
%       reddening. The brown line shows an approximation of the continuum.   
%              }
%       \vspace*{-3mm} 
%         \label{pg0043_hst.ps}
%   \end{figure*}
%
Overlaid is a second UV spectrum of PG0043+03 that
 was taken before with the
HST-FOS spectrograph in 1991
(Turnshek et al.\citealt{turnshek94}). This spectrum 
covered the intrinsic wavelength range from 
$\sim$1130\,$\AA\ $ to $\sim$2300~$\AA\ $.
The continuum and line fluxes increased by a factor of $\sim$1.8
between the years 1991 and 2013
when comparing the overlapping wavelength range
of both UV spectra
(Kollatschny et al.\citealt{kollatschny15b}).
%Fig.~\ref{pg0043_cyc_lambda.ps} shows combined ultraviolet spectrum
%of PG0043+039.
We multiplied
 the observed UV spectrum taken in 1991 with this factor of 1.8
(in Fig.\,\ref{pg0043_cyc_lambda.ps})
to match the UV observations of 2013.
The HST spectra
have been smoothed ($\Delta \lambda = 5.7\,\AA$ or $8\,\AA$)
 for high-lightening
weaker spectral structures. 
We corrected the UV, as well as our optical spectra of PG0043+039
for a Galactic extinction E(B-V) = 0.02087 that was taken from
Schlafly \& Finkbeiner
\cite{schlafly11}.
Overlaid is a power-law continuum F$_\lambda$~$\sim$~$\lambda^{\alpha}$ with
$\alpha = 0.67 \pm 0.02$ (in Fig.~\ref{pg0043_cyc_lambda.ps}).
For fitting this UV continuum, 
we used the spectral windows at 895, 983, 1360, 1609, and 1689 $\AA$ with
typical widths of 10 $\AA${}.

\begin{figure}
\hbox{
    \includegraphics[width=4.3cm,angle=0]{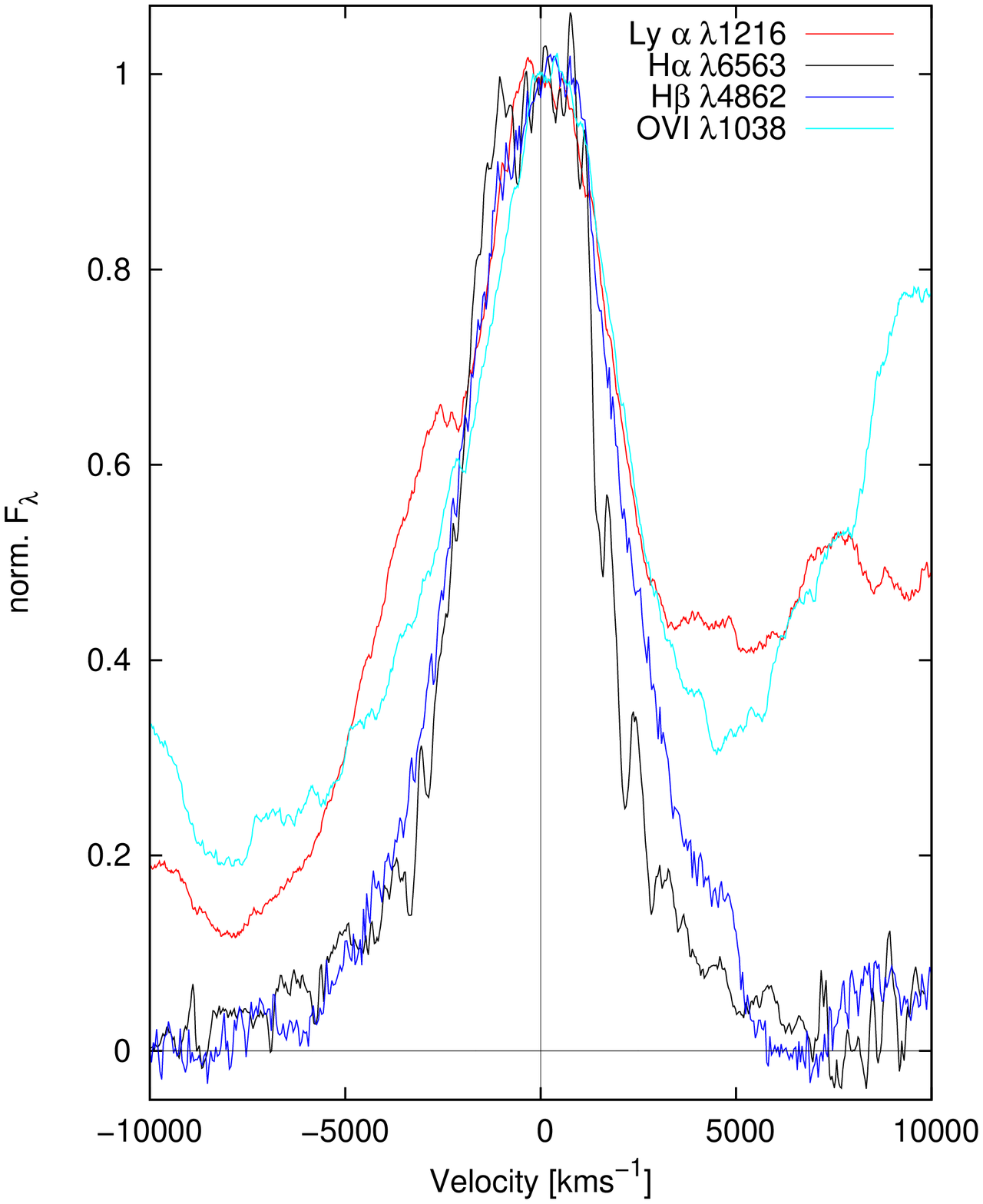}
    \includegraphics[width=4.3cm,angle=0]{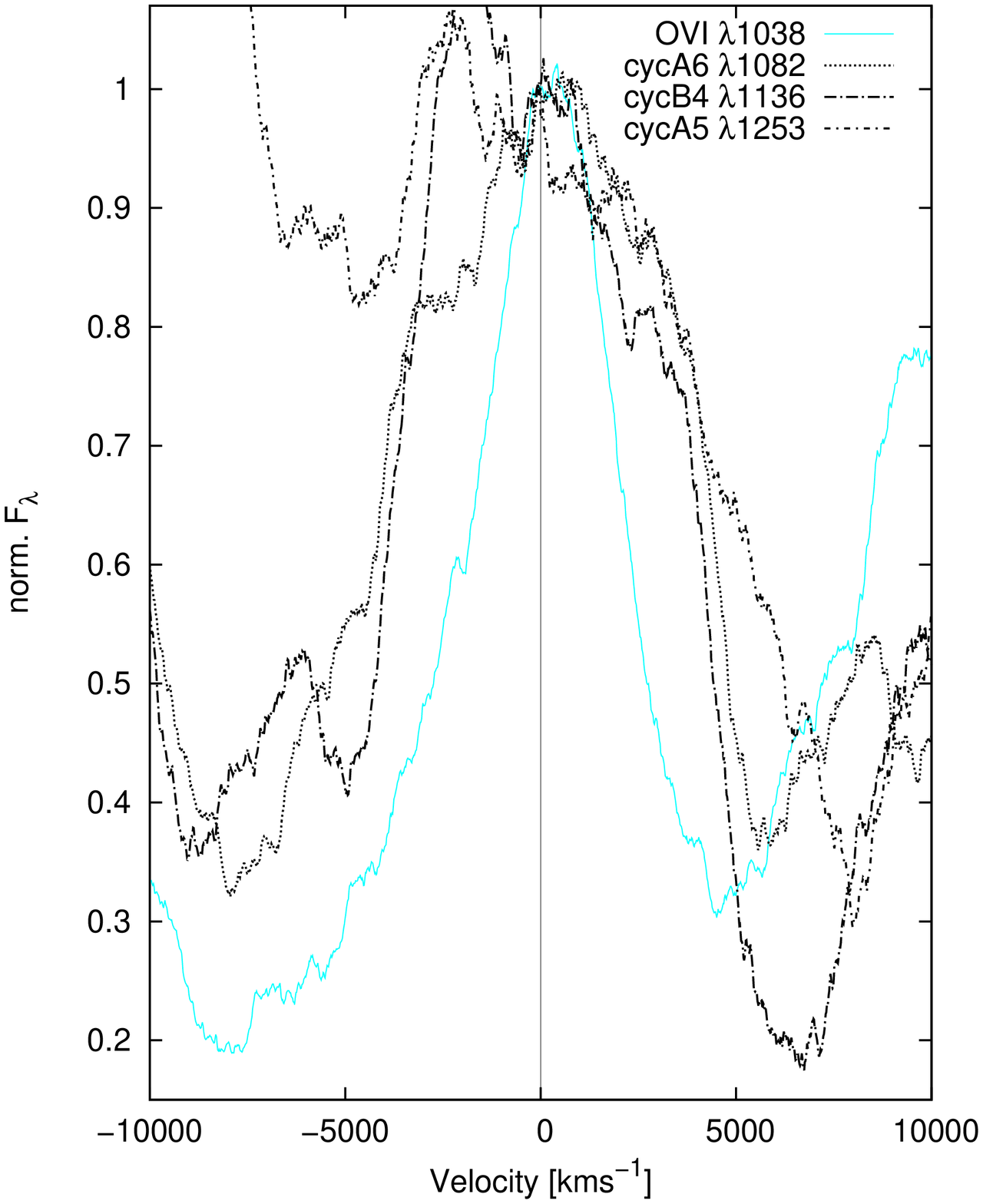}
}
      \caption{Left: Normalized emission line profiles of the
 strongest optical and UV emission lines in velocity space.
Right: Cyclotron emission line profiles of CycA2, CycB2,
and CycA3 at $\lambda 1082$, $\lambda 1136$, and $\lambda 1253$ 
in comparison to
% the normalized emission line profiles of
the \ion{O}{vi}\,$\lambda 1038$ line.} 
% The Ly$\alpha$ line
% has been shifted by 855 [\kms] to the red for comparing the profiles.} 
      \vspace*{-4mm} 
         \label{pg0043_velo_profile_lya855_cont2.ps}
   \end{figure}
   \begin{figure}
    \includegraphics[width=6.3cm,angle=270]{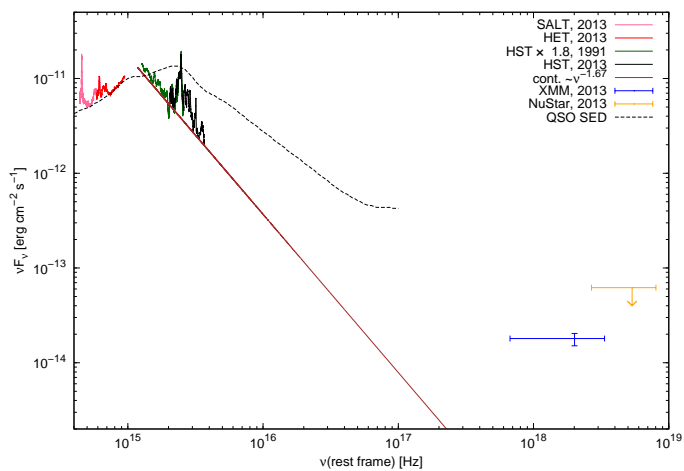}
      \caption{Observed multiwavelength spectral distribution of PG0043+039
from the optical to the X-ray ranges in July 2013. An extrapolation
of the UV continuum gradient to the X-ray regime is indicated.
%dashed line from Tab. 3  column 'ALL' of Richards et al. 2011 (AJ 141,167)
%     taken in 1990/1991 and 2013.
The dashed line indicates the spectral energy distribution
from the optical to the X-ray 
 for a typical AGN.
              }
       \vspace*{-3mm} 
         \label{pg0043_uvoptx2.ps}
   \end{figure}

There is no detectable Lyman edge associated with the BAL absorbing gas.
Furthermore, we indicate the identifications
of the strongest UV emission lines Ly$\alpha$, $\ion{O}{vi}\,\lambda 1038$,
etc., of the geo-coronal lines, 
as well as of other emission lines that we attribute
to two cyclotron systems A and B with
their second to seventh harmonics. The integer numbers identify the
emissions with multiples of the cyclotron fundamental.
Longward of 1600 $\AA$, the spectrum is contaminated by FeII and FeIII blends.
The observed UV/far-UV spectrum of PG0043+039 is quite different
from
the UV spectra of other quasars or BALQs
(Hall et al. \citealt{hall02}, Baskin et al. \citealt{baskin13}, Saez et al.
\citealt{saez12}).

The strongest atomic emission lines in the UV are the
Ly$\alpha$ and \ion{O}{vi}\,$\lambda 1038$ lines.
The strongest emission lines in the optical
spectra of PG0043+039 are the Balmer lines H$\alpha$ and H$\beta$.
We show the optical and UV profiles of these lines jointly in
Fig.\,\ref{pg0043_velo_profile_lya855_cont2.ps} in velocity space.
They are normalized to the same maximum intensity.
All these broad emission lines exhibit very similar profiles and
 have nearly identical line
widths of 4000 to 4800 \kms (FWHM) except Ly$\alpha$, which shows a
slightly broader line
width of 6300 \kms. This difference might be simulated by a different
determination of the underlying continuum caused
by additional underlying line components.
%or due to an 
%additional underlying component in the Ly$\alpha$ line. 
%wk: red and blueshift of Lya OVI
%profile and additional underlying components in Ly$\alpha$ line
%i.e. $\ion{Si}{iii}\,\lambda 1206$ (see Fig.~\ref{pg0043_hst_2013_1991.ps})
Furthermore, we present the profiles
of the strongest
UV humps that we attribute to the cyclotron lines CycA6, CycB4,
and CycA5 -- in velocity space -- in comparison to the
\ion{O}{vi}\,$\lambda 1038$ line.
Again they are normalized to the same maximum intensity. All the 
cyclotron lines show similar line widths of about 10\,000 \kms,
on the one hand, and as a whole, different line shapes
than those of the normal emission lines,
% (FWHM$\sim5000~$\kms)
on the other hand.
  
Figure\,\ref{pg0043_uvoptx2.ps} presents the observed optical-UV-X-ray
spectral distribution of PG0043+039 in July 2013.
Given are the optical 
(taken with the SALT and HET telescopes), the UV, and the far-UV spectra taken with
the HST, the X-ray flux taken with the XMM-Newton satellite, as well as an upper
limit in the hard X-ray range (8--24 keV) obtained with NuSTAR
(Luo et al.\citealt{luo14}). The UV spectrum taken in 1991 has been multiplied
with a factor of 1.8 to match the far-UV spectrum in 2013.  
The extrapolation of the UV  
 power-law continuum F$_\nu$~$\sim$~$\nu^{\alpha}$ with
$\alpha = -2.67 \pm 0.02.$
%(with  $\alpha_{ox}=-2.3 \pm 0.3$) to the X-ray
is in good agreement with the extreme
faint X-ray flux  $\alpha_{ox}=-2.37 \pm 0.05$.
The figure shows the mean
composite SED (derived from
a sample of 259 quasars, Richards et al.\citealt{richards06}) for comparison.
We scaled this composite spectrum at the frequency of $10^{15}$ Hz
with respect to the spectrum of PG0043+039. 
The Balmer lines are the strongest optical emission lines
in the
spectrum of PG0043+039 besides their strong \ion{Fe}{ii} blends
(Turnshek et al.\citealt{turnshek94}, Kollatschny et
al.\citealt{kollatschny15b}).
%In general, the optical spectrum is similar to that of Mrk\,231. However, the 
%FeII blends are even stronger in  Mrk\,231 
%(Boksenberg et al. \citealt{boksenberg77},
%Lipari et al. \citealt{lipari09}) in comparison to PG0043+039.
%
The strongest atomic emission lines of AGNs
in the UV/far-UV range of 800 -- 1500 \AA\  are the
Ly$\alpha$, \ion{O}{vi}\,$\lambda 1038$, and \ion{N}{v}\,$\lambda 1243$ lines.
%(see  Fig.~\ref{pg0043_cyc_lambda.ps}).
%(see  Fig.~\ref{pg0043_hst.ps}).
We observe very similar line ratios of these UV emission lines 
%the \ion{O}{vi}\,$\lambda 1038$/Ly$\alpha$/\ion{N}{v}\,$\lambda 1243$   
in comparison to those seen in a mean composite spectrum
(Shull et al.\citealt{shull12}). However, PG0043+039 shows additional
strong broad lines or humps in the UV spectrum  
that could not be attributed to known emission lines
based on their wavelength positions and relative intensities.
%, and line widths.
We compared the UV spectrum of PG0043+039 with those of
 IZw1 (Laor et al.
\citealt{laor97}), with a composite quasar spectrum (Vanden Berk et al.
\citealt{vandenberk01}), with Mrk\,231 (Veilleux et al.\citealt{veilleux13},
 Lipari et al. \citealt{lipari09}), and with  
ultraviolet composite spectra of AGN based on HST-COS observations
(Shull et al.\citealt{shull12}, Stevans et al.\citealt{stevens14}).
Especially the strong broad emission humps/lines at 
1082, 1136, and 1253 \AA\
 could not be attributed in a simple way
to known UV emission lines based on their wavelengths and relative intensities. 
Before we demonstrate an alternative solution in the next section
that these lines can be attributed to cyclotron lines
we discuss the presence of line emissions
at these wavelength ranges in other AGNs first:\\
- The hump at 1082 \AA\ is not seen in the composite AGN spectra of 
Shull et al.\cite{shull12} and Stevans et al. \cite{stevens14}.
They list an expected \ion{N}{ii} line at $\lambda 1085$. However, there
is an unidentified emission feature in the composite AGN spectra that can 
 only be seen  shortward of this position.
Furthermore, these authors assume
a continuum window to be there in the
wavelength range 1080--1110 \AA{}. In the specific case
of the quasar HS\,1103+3441 (TON\,1329), an emission bump seems
to be present at around 1082 \AA\ that could be attributed to
the \ion{N}{ii} line at $\lambda 1085$. However, this galaxy
shows an extreme [\ion{N}{iii}] $\lambda 3968$ line in the optical 
spectrum taken with Sloan Digital Sky Survey
(Schneider et al., \citealt{schneider07}).
This optical line is not present in the spectrum of PG0043+039.\\  
- At 1136 \AA\ no spectral feature is to be seen in the composite AGN spectra.
On the other hand, Stevans et al. \cite{stevens14}
find a weak spectral line of \ion{Fe}{iii}  at
$\lambda 1126$ \AA{}. This line might contribute
to the outermost blue edge in our 1136 \AA\  emission hump. Veilleux et al.
\cite{veilleux13} identify a very weak feature at $\lambda 1133$ \AA\
in Mrk\,231. They claim that an identification of this feature
with lines of NI or FeII is unlikely.
No emission line is to be seen at 1253 \AA,\ in either Mrk\,231 or the composite AGN spectra.\\
In principle, there is a weak possibility that the observed humps might be
caused by low ionization lines or blends of FeII or NII, for example. However, in that
case one should be able to explain why
all these emission features have the same widths,  
broader than those of the regular emission lines. 
Additionally, these bumps clearly protrude out of the continuum: they
are no normal line blends.
 Mrk\,231 shows weak unidentified humps in the UV as well 
(Veilleux et al.\citealt{veilleux13}), but at different wavelength
positions and at fainter intensities relative to PG0043+039.
%Furthermore,  the optical \ion{Fe}{ii} line blends are stronger in Mrk\,231
%than in PG0043+039.
%
% 
%

\subsection{Cyclotron emission lines in PG0043+039}

Besides known UV emission
lines, such as as Ly$\alpha$  and \ion{O}{vi}\,$\lambda 1038,$ 
other broad line humps are to be seen in the UV spectrum of PG0043+039.
%These broad line humps have different/broader
%line widths than the atomic emission lines.
% Furthermore their wavelength
%positions are different relative to known UV emission lines of other AGN
%(e.g. Shull et al., \citealt{shull12},  Veilleux et al. \citealt{veilleux13}).
%as well as their relative line intensities.
One possible explanation for these lines it that we
assigned these emission lines to
cyclotron emission lines.  
A modeling of the observed features with cyclotron emission lines 
should simultaneously explain the wavelength positions, the relative distances
between the lines,  the relative intensities of the lines, and
the widths of the lines.

Cyclotron emission lines have been established so far in
 the UV, optical, and infrared (IR) spectra
of AM Herculis stars - so-called polars hosting strongly magnetic
white dwarfs, as well as
in intermediate polars. Typical magnetic field strengths are on the order of
B = $3-15\times10^{7}$ G in the inner magnetic accretion regions.
Examples of observed and modeled cyclotron lines in AM Her stars are
shown, for example, in 
 Schwope et at.
\cite{schwope06},
 Campbell et al. \cite{campbell08}, and references therein.  
%In polars the strong magnetic field of the white dwarf suppresses the formation
%of an accretion disk.
%In intermediate polar stars the magnetic field of the
%white dwarf is not strong enough to entirely control the accretion flow.
%Here only a truncated accretion disk
%can form at its interior of the magnetospheric radius.
 In the standard model for accretion onto a magnetic white dwarf
(e.g., Lamb and Masters \citealt{lamb79})
%, Burwitz \citealt{burwitz97})
an adiabatic
standing shock forms in the accretion column above the surface of the white
dwarf at highly supersonic speeds.
In the shock region, the kinetic energy is transformed into thermal
energy, and the matter is slowed down into a subsonic settling flow. During
this process the matter in the settling flow is heated to a
shock temperature of $10^{8} - 10^{9}$ K.
%The height of the shock above the surface
%depends on the accretion rates and the efficiency of the cooling mechanisms.
The hot matter in the settling flow is then cooled down by thermal
bremsstrahlung
% in the hard X-ray regime
and/or cyclotron radiation in the UV-to-near-IR wavelength range.  
%Further cyclotron lines have been detected in the hard X-ray spectra
%in a handful of neutron stars
%(e.g. Tr\"umper et al., \citealt{truemper78} or Wilms  \citealt{wilms14} for a %recent review).
%Neutron stars show magnetic field strengths of the order of $10^{12}$ G.
Cyclotron lines are emitted by nonrelativistic electrons in strong magnetic
fields. Cyclotron emission occurs at the fundamental frequency 
${\omega}_{cyc}$ that is given by
$\omega_{cyc} = e \times B~/~m{_e} \times c$
and its higher harmonics n$\omega_{cyc}$, where B is the
magnetic field strength, and e and $m_e$ are charge and mass of the electron
(e.g., Fischer \& Beuermann \citealt{fischer01}, and references therein). 
%
%\begin{equation}
%  \label{eq:cyclotron}
%  \omega_{cyc} = \frac{eB}{\mathrm{m_{e}c}} = 2\pi{}c\left(\frac{1}{\lambda_{n+%1}}-\frac{1}{\lambda_{n}}\right) \quad\quad n= 1,2,3,\dots
%\end{equation}
%
A broadening of
the cyclotron spikes is caused, among others, by the high temperatures 
of the plasma and/or orientation or gradients in the magnetic field.

%We made various tests whether the humps seen in the UV spectrum of PG0043+039
%can be attributed to cyclotron emission lines.
% The existence of 
%magnetohydrodynamic standing shocks in in-flowing plasmas around black holes  
%has been studied before by e.g. Fukumura et al \cite{fukumura07} and references
%therein.
%This shock formation can be a plausible mechanism for creating 
%hot and/or strongly magnetized plasma regions which could be associated
%with subsequent non-thermal emission as well. 
Magnetohydrodynamical shock formation is possible, for example,
in equatorial/non-equatorial
plasma flows close to the black-hole event horizon
(Fukumura et al.\citealt{fukumura07}, and references therein).
Shocks in these MHD plasmas might be
responsible for creating very hot T $\approx \times10^{9}$ K and/or
strongly magnetized plasma regions.
These shocks could be the origin of cyclotron radiation
similar to the origin of cyclotron radiation connected to shocks in CV stars. 

We carried out various tests of whether it is
possible to model the observed humps in the UV spectrum of PG0043+039
 with cyclotron emission lines and their harmonics.
We used a program described by Fischer \& Beuermann
\cite{fischer01} originally developed for cyclotron radiation emitted from
standing shocks above accreting magnetic white dwarfs.
%The modeling should explain simultaneously the relative distances
%between the lines, the relative intensities of the lines, as well as
%the widths of the lines.
The physical parameters that affect the pattern of the cyclotron emission
 lines 
are the magnetic field strength B of the line emitting region, as well
as their temperature T. 
An additional argument for modeling the lines is the dimensionless parameter
$\Lambda$ with $\Lambda=4\pi$ e n$_{e}$ l / B, where n$_{e}$ is the electron
density and l is the size of the line emitting region.
%the line of sight view with respect to the magnetic field lines. 
At the end we needed two
cyclotron line systems that we called A and B systems
for modeling the UV emission humps in PG0043+039.
% blue (B) and red (A) systems.
We show in Fig.~\ref{pg0043_cyc_lambda.ps}
% ~\ref{pg0043_cyc_nu.ps}, ~\ref{pg0043_cyc_lambda.ps}
the computed wavelength position of the two cyclotron systems (A, B)  with their
second to seventh harmonics.
% in frequency and lambda space.
%The integer numbers identify the
%emission humps with multiple of the cyclotron fundamental.

Our best fit to the observations yields plasma
temperatures of T = 3.8~keV (A) (T $\approx 4 \times10^{7}$ K) and
 T = 1.9~keV (B) (T $\approx 2 \times10^{7}$ K)
and field strengths
% in the accretion regions
of B = 1.95 (A) and 1.45 (B) $\times10^{8}$ G
 and log\,$\Lambda$ values of 4 and 7. This corresponds to
line emitting region sizes of $10^{15}$ - $10^{18}$ cm$^{2}$ for assumed
density values of n$_{e}~=10^{15}\,cm^{-3}$.
 We did not correct for gravitational redshift
or other relativistic effects.
Since there is no theory for the formation of
cyclotron lines originating in hot and highly magnetized plasma
in the vicinity of super-massive black holes our solution is  only
a first-order approximation.
On the other hand, it gives the first direct
measurement of the magnetic field next to a black hole. 
The two observed systems might be explained by line emission from both 
magnetic poles.
Two poles accretion configurations are known in a few cases
for AM Her stars as well (e.g., Schwope et al. \citealt{schwope99}).
The underlying broad absorption feature of the  $\ion{C}{iv}\,\lambda 1548$ line
might be responsible for the overestimated cyclotron lines intensities
at 1419 and 1521 \AA\ in Fig.\,\ref{pg0043_cyc_lambda.ps}.
A final confirmation of the presence of cyclotron lines 
in the UV spectra of AGN could be achieved through spectropolarimetry. 

%Magnetic fields with field strengths of the order of  $10^{8}$ G
%are possible within the models of magnetically-dominated accretion flows
%(MDAF) into black holes (Meier \citealt{meier12} and references therein).
%These magnetic field
%strengths are scaled with respect to the central black hole masses as well
%as their radii. The MDAF regions are located interior to a transitional flow
%region and an accretion-dominated accretion flow region (ADAF). 

\section{Summary}

The extreme X-ray faint quasar PG0043+039 has finally been detected in
a deep X-ray exposure with XMM-Newton as part of a multifrequency campaign.
 The far-UV spectrum, which was taken with the HST,
turned out to be very peculiar in terms of
showing various broad humps in addition
to the normal emission lines.
A modeling of these observed emission humps with cyclotron lines can explain
the wavelength positions and their relative distances, 
 as well as their relative intensities.
We derived  plasma
temperatures of T $\sim$ 3 ~keV
and magnetic field strengths of  B = 2 $\times10^{8}$ G
for the line emitting regions close to the black hole.

\begin{acknowledgements}
This paper is based on observations taken with the XMM-Newton, HST, SALT,
and HET Telescopes. We thank Klaus Reinsch and Axel Schwope for their
advice on cyclotron line emission. We thank David Meier
for valuable suggestions.
This work has been supported by DFG grant Ko 857/32-2.
%
%We deeply thank A. Talavera for providing the XMM-Newton OM V-grism spectrum of%  PG 0043+039.

\end{acknowledgements}


\begin{thebibliography}{}
%
%\input{references}
%
%\bibitem[\protect\citeauthoryear{} {1996}]{Arnaud1996} 
%        Arnaud, K. A., 1996, ASPC, 101, 17
%
% \bibitem[\protect\citeauthoryear{} {2007}]{atlee07} Atlee, D. W., Gould, A.
%     2007, ApJ, 664, 53
%
%\bibitem[\protect\citeauthoryear{} {1993}]{bahcall93} Bahcall, J. N.,
%     et al. 1993, ApJS, 87, 1
%
 \bibitem[\protect\citeauthoryear{} {2013}]{baskin13} Baskin, A.,
   Laor, A., \& Hamann, F.  2013, MNRAS, 432, 1525 
% 
%\bibitem[\protect\citeauthoryear{} {2002}]{bechtold02} Bechtold, J.,
%     Dobrzycki, A., Wilden, B., et al. 2002, ApJS, 140, 143
%
%\bibitem[\protect\citeauthoryear{} {2009}]{bianchi09} Bianchi, S.,
%     Pinconcelli, E., Chiaberge, M., et al. 2009, ApJ, 695, 781
%
 \bibitem[\protect\citeauthoryear{} {1977}]{boksenberg77} Boksenberg, A.,
  Carswell, R. F., Allen, D. A., et al. 1977, MNRAS, 178, 451 
%
% \bibitem[\protect\citeauthoryear{} {1992}]{boroson92} Boroson, T. A.,
%   Green, R. F. 1992, ApJS, 80, 109
%
% \bibitem[\protect\citeauthoryear{} {1997}]{burwitz97} Burwitz, V.,
%   1997, PhD thesis G\"ottingen University  on 'X-ray and optical
%    properties of magnetic CVs'
%
 \bibitem[\protect\citeauthoryear{} {2000}]{brandt00} Brandt, W. N.,
   Espey, B. R., Kopko, JR., M., et al. 2000, ApJ, 528, 637 
%
 \bibitem[\protect\citeauthoryear{} {2008}]{campbell08} Campbell, R. K.,
   Harrison, T. E., Schwope, A. D. et al..  2008, ApJ, 672, 531 
%
 \bibitem[\protect\citeauthoryear{} {2008}]{czerny08} Czerny, B.,
   Siemiginowska, A., Janiuk, A. et al..  2008, MNRAS, 386, 1557 
%
%\bibitem[\protect\citeauthoryear{} {2011}]{fabian11} Fabian, A. C.,
%  Zoghbi, A., Wilkins, D., et al. 2011, MNRAS, 419, 116 
%
 \bibitem[\protect\citeauthoryear{} {2001}]{fischer01}
  Fischer, A., \& Beuermann, K. 2001,   A\&A, 373, 211
%
% \bibitem[\protect\citeauthoryear{} {1999}]{fitzpatrick99}     
%       Fitzpatrick, E. L. 1999, PASP, 111, 63 
%
 \bibitem[\protect\citeauthoryear{} {2007}]{fukumura07} Fukumura, K.,
    Takahashi, M., Tsuruta, S.  2007, ApJ, 657, 415 
%
% \bibitem[\protect\citeauthoryear{} {1999}]{gallagher99} Gallagher, S. C.,
%   Brandt, W. N., Sambruna, R., M., et al. 1999, ApJ, 519, 549 
%
%\bibitem[\protect\citeauthoryear{} {2009}]{gibson09} Gibson, R. R.,
%   Jiang, L., Brandt, W. N., et al. 2009, ApJ, 692,  
%
%\bibitem[\protect\citeauthoryear{} {1991}]{haardt91} Haardt, F,
%   \& Maraschi, L. 1991, ApJ, 380, L51
%
\bibitem[\protect\citeauthoryear{} {2002}]{hall02} Hall, P. B.,
   Anderson, S. F., Strauss, M. A. et al. 2002, ApJS, 141, 267
%
%\bibitem[\protect\citeauthoryear{} {2013}]{hamann13} Hamann, F.,
%  Chartas, G., McGraw, S., et al. 2013, MNRAS, 435, 133 
%
% \bibitem[\protect\citeauthoryear{} {2009}]{ho09} Ho, L. C.,
%   Kim, M. 2009, ApJS, 184, 398
%
% \bibitem[\protect\citeauthoryear{} {2001}]{Jansen2001}
%  Jansen, F., Lumb, D., Altieri, B., et al. 2001,   A\&A, 365, L1
%
% \bibitem[\protect\citeauthoryear{} {1989}]{kellermann89} Kellermann, K., I.
%  et al. 1989, AJ, 98, 1195 
%
% \bibitem[\protect\citeauthoryear{} {2006}]{koide06} Koide, S., Kudoh, T., 
%    Shibata, K.   2006, Phys.Rev.D. 74, 044005  
%
% \bibitem[\protect\citeauthoryear{} {2003a}]{kollatschny03a} Kollatschny, W.
%   2003a, A\&A, 407, 467 
%
% \bibitem[\protect\citeauthoryear{} {2003b}]{kollatschny03b} Kollatschny, W.
%   2003b, A\&A, 412, L61 
%
  \bibitem[\protect\citeauthoryear{} {2001}]{kollatschny01} Kollatschny, W.,
   Bischoff, K., Robinson, E. L., et al. 2001, A\&A, 379, 125 
%
%\bibitem[\protect\citeauthoryear{} {1992}]{kollatschny92} Kollatschny, W.,
%   Dietrich, M., Hagen, H. 1992, A\&A, 264, L5 
%
%\bibitem[\protect\citeauthoryear{} {1985}]{kollatschny85} Kollatschny, W.,
%   Fricke, K. 1985, A\&A, 146, L11 
%
 \bibitem[\protect\citeauthoryear{} {2015b}]{kollatschny15b} Kollatschny, W.,
   Schartel, N., Zetzl, M., et al. 2015b, A\&A, in prep.
%
\bibitem[\protect\citeauthoryear{} {1979}]{lamb79}
  Lamb, D. Q., Masters, A. R. 1979, ApJ, 234, 117 
%
\bibitem[\protect\citeauthoryear{} {1997}]{laor97}
  Laor, A., Jannuzi, B. T., Green, R. F., et al. 1997, ApJ, 489, 656 
%

%\bibitem[\protect\citeauthoryear{} {2014}]{laor14}
%  Laor, A. \& Davis, S. W., 2014, MNRAS, 438, 3024 
%
%\bibitem[\protect\citeauthoryear{} {2009}]{leighly09}
%  Leighly, K. M., Hamann, F., Casebeer, D. A., et al. 2009, ApJ, 701, L176 
%
\bibitem[\protect\citeauthoryear{} {2014}]{luo14} Luo, B., Brandt, W. N.,
 Alexander, D. M., et al. 2014, ApJ, 794, 70
%
 \bibitem[\protect\citeauthoryear{} {2009}]{lipari09} Lipari, S.,
  Sanchez, S. F., Bergmann, M., et al. 2009, MNRAS, 392, 1295 
%
% \bibitem[\protect\citeauthoryear{} {2001}]{Mason2001} 
%Mason, K. O., Breeveld, A., Much, R., Carter, M., et al., 2001, A\&A, 365, L36
%
%\bibitem[\protect\citeauthoryear{} {2012}]{meier12} Meier, D.
%  2012, 'Black Hole Astrophysics', Springer Link 
%
%\bibitem[\protect\citeauthoryear{}{2007}]{Nandra2007}
%Nandra, K., O'Neill, P. M., George, I. M. \& Reeves, J. N.,
%2007, MNRAS, 382, 194
%
%\bibitem[\protect\citeauthoryear{} {2012}]{meusinger12} Meusinger, H.,
%   Schalldach, B., Scholz, R.-D., et al. 2012, A\&A, 541, A77
%
%\bibitem[\protect\citeauthoryear{}{2012}]{Page2012}
%Page, M. J., Brindle, C., Talavera, A., et al., 2012, MNRAS, 426, 903
%
 \bibitem[\protect\citeauthoryear{} {2005}]{Piconcelli2005} Piconcelli, A.,
 Jim\'{e}nez-Bail\'{o}n, E., Guanazzi,M. et al. 2005,
A\&A, 432, 15
%
\bibitem[\protect\citeauthoryear{} {2006}]{richards06}
  Richards, G. T., et al., 2006, ApJS, 166, 470 
%
%\bibitem[\protect\citeauthoryear{} {2005}]{risaliti05}
%  Risaliti, G., Elvis, M., Fabbiano, G., et al., 2005, ApJ, 623, L93 
%
\bibitem[\protect\citeauthoryear{} {2012}]{saez12}
  Saez, C., Brandt, W. N., Gallagher, S. C., et al., 2012, ApJ, 759, 42 
%
%\bibitem[\protect\citeauthoryear{} {1993}]{Savage1993} 
%Savage, B. D., Lu, L., Bahcall, J. N., et al., 1993, ApJ, 413, 116
%
 \bibitem[\protect\citeauthoryear{} {2007}]{Schartel2007}
  Schartel, N., Rodr\'{i}guez-Pascual, P.M.,  Santos-Lle\'{o}, M., et al.
2007,  A\&A, 474, 431
%
%\bibitem[\protect\citeauthoryear{} {2010}]{schartel10}
%  Schartel, N., Rodr\'{i}guez-Pascual, P.M.,  Santos-Lle\'{o}, M., et al.
%2010,  A\&A, 512, 75 
%
  \bibitem[\protect\citeauthoryear{} {2011}]{schlafly11} Schlafly, E. F., 
   \& Finkbeiner D. P. 2011, ApJ., 737, 103
%
% \bibitem[\protect\citeauthoryear{} {1998}]{schlegel98} Schlegel, D. J.,
% Finkbeiner, D. P., \& Davis, M. 1998, ApJ, 500, 525
%
 \bibitem[\protect\citeauthoryear{} {1983}]{schmidt83} Schmidt, M.,
   Green, R. F. 1983, ApJ, 269, 352
%
 \bibitem[\protect\citeauthoryear{} {2007}]{schneider07} Schneider, D. P.,
   et al. 2007, AJ, 134, 102
%
 \bibitem[\protect\citeauthoryear{} {1999}]{schwope99}
   Schwope, A. D., Schwarz, R., \& Greiner, J. 1999, A\&A, 348, 861
%
 \bibitem[\protect\citeauthoryear{} {2006}]{schwope06}
   Schwope, A. D., Schreiber, M. R., \& Szkody, P. 2006, A\&A, 452, 955
%
% \bibitem[\protect\citeauthoryear{} {2014}]{scott14} Scott, A. E., 
%   \& Stewart G. C. 2014, MNRAS, 438, 2253
%
%\bibitem[\protect\citeauthoryear{} {2009}]{serjeant09}
%Serjeant, S., \& Hatziminaoglou, E., 2009 MNRAS, 397, 265 
%
% \bibitem[\protect\citeauthoryear{} {2014}]{shappee14} Shappee, B. J., 
%   Prieto, J. L., Grupe, D. et al. 2014, ApJ., 788, 48
%
\bibitem[\protect\citeauthoryear{} {2012}]{shull12}
 Shull, J. M., Stevans, M., \& Danforth, C. W. 2012, ApJ, 752, 162
%
% \bibitem[\protect\citeauthoryear{} {2006}]{steffen06}
%   Steffen, A. T., Strateva, I., Brandt, W. N., et al. 2006, AJ, 131, 2816
%
\bibitem[\protect\citeauthoryear{} {2014}]{stevens14}
  Stevans, M., Shull, J. M., \& Danforth, C. W., et al. 2014, ApJ, 794, 75
%
% \bibitem[\protect\citeauthoryear{} {2001}]{Strueder2001}
%   Str\"uder, L., Briel, U., Dennerl, K., et al. 2001, A\&A, 365, L18
%
% \bibitem[\protect\citeauthoryear{} {2000}]{sulentic00}
%   Sulentic, J. W., Marziani, P., Dultzin-Hacyan, D. 2000, ARA\&A, 38, 521
%
% \bibitem[\protect\citeauthoryear{} {1978}]{truemper78} Tr\"umper, J.,
%   Pietsch, W., Reppin, C. et al. 1978, ApJ, 219, L105 
%
% \bibitem[\protect\citeauthoryear{} {2001}]{Turner2001}
%  Turner,  M.J.L., Abbey, A., Arnaud, M.,  et al. 2001, A\&A, 365, L27
%
 \bibitem[\protect\citeauthoryear{} {1994}]{turnshek94} Turnshek, D. A.,
   Espey, B. R., Kopko, JR., M., et al. 1994, ApJ, 428, 93 
%
% \bibitem[\protect\citeauthoryear{} {1997}]{turnshek97} Turnshek, D. A.,
%   Monier, E. M., Sirola, C. J., et al. 1997, ApJ, 476, 40 
%
 \bibitem[\protect\citeauthoryear{} {2001}]{vandenberk01} Vanden Berk, D. E.,
   Richards, G. T., Bauer, A., et al. 2001, ApJ, 122, 549 
%
 \bibitem[\protect\citeauthoryear{} {2013}]{veilleux13} Veilleux, S.,
   Trippe, M., Hamann, F., et al. 2013, ApJ, 764, 15 
%
%\bibitem[\protect\citeauthoryear{} {1996}]{Verner1996} 
%Verner, D. A., Ferland, G. J., Korista, K. T. \& Yakovlev, D. G., 1996, ApJ, 46%5, 487
%
% \bibitem[\protect\citeauthoryear{} {2001}]{Watson2001}
%  Watson, M. G., Augu\'eres, J.-L., Ballet, J., et al. 2001,  A\&A, 365, L51
%
% \bibitem[\protect\citeauthoryear{} {2009}]{Watson2009}
%  Watson, M. G., Schr{\"o}der, A. C., Fyfe, D., et al. 2009,  A\&A, 493, 339
%
% \bibitem[\protect\citeauthoryear{} {2000}]{Wilms2000} 
%Wilms, J., Allen, A. \& McCray, R., 2000,  ApJ, 542, 914,
%
% \bibitem[\protect\citeauthoryear{} {2014}]{wilms14} 
%Wilms, J. 2014,  EPJ Web of Conferences, Vol. 64, id.06001
%
\end{thebibliography}
\end{document}